\documentclass{svproc}
\usepackage{xcolor}
\usepackage{amsmath,amssymb}
\usepackage{hyperref}
\usepackage{epsfig}
\usepackage{amssymb}
\usepackage{graphicx,float}
\usepackage{subcaption}
\usepackage{url}

\date{}
\begin{document}

\mainmatter

\title{Small Number of Communities in Twitter Keyword Networks}

\author{Linda Abraham\inst{1} \and Anthony Bonato\inst{1} \and Alexander Nazareth\inst{1}}

\institute{Ryerson University, Toronto, Ontario, Canada}

\maketitle

\begin{abstract}
We investigate networks formed by keywords in tweets and study their community structure. Based on datasets of tweets mined from over seven hundred political figures in the U.S.\ and Canada, we hypothesize that such Twitter keyword networks exhibit a small number of communities. Our results are further reinforced by considering via so-called pseudo-tweets generated randomly and using AI-based language generation software. We speculate as to the possible origins of the small community hypothesis and further attempts at validating it.
\end{abstract}

\section{Introduction}\label{intro}

Twitter is a dominant social media and micro-blogging platform, allowing users to present their views in concise 280-character tweets. An active social media presence has become the mainstay of modern political discourse in the United States and Canada; many politicians, such as members of Congress and members of Parliament, frequently tweet. The corpus of tweets by such political figures forms a massive data source of regular updates on government strategy and messaging. Tweets may reveal approaches to reinforce political platforms, describe policy, or either bolster support from followers or antagonize political adversaries. Besides their political content, the mining and analysis of tweets by political figures may lead to fresh insights into the structure and evolution of networks formed by Twitter keywords.

In \emph{Twitter keyword networks}, the nodes are keywords, which are significant words, distinguished from common stop words such as ``and'' or ``the.'' Nodes are adjacent if they are in the same tweet; we may consider this a weighted graph, where multiple edges arise from multiple occurrences of keyword pairs. These are \emph{co-occurrence networks} of keywords in tweets, and the extraction and analysis of co-occurrence networks provide a quantitative method in the large-scale analysis of such tweets. Networked data may be mined from Twitter, and algorithms applied to probe the community structure of the resulting networks.

\begin{figure}[!h]
  \centering
  \includegraphics[width=9cm]{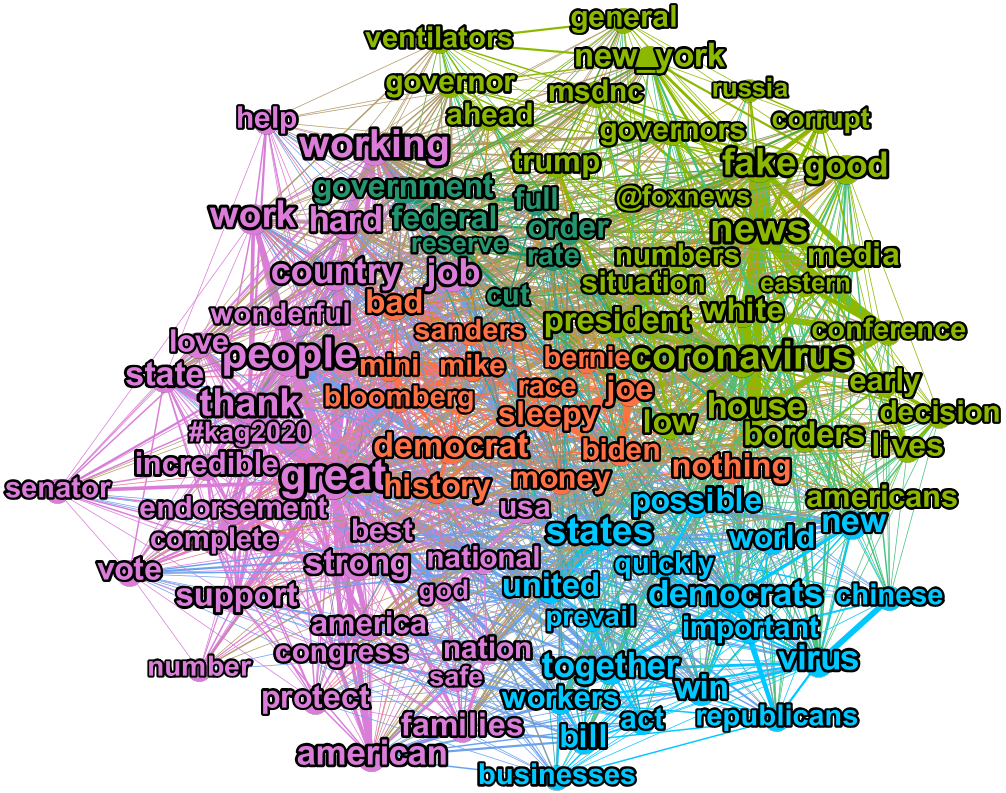}
  \caption{Keyword network of tweets by President Trump from March of 2020.} \label{trump}
\end{figure}

See Figure~\ref{trump} for an example of the approach described in the previous paragraph, taken from March 2020 of tweets by then-President Donald J.\ Trump. We chose this month as it was the beginning of major lockdowns owing to the COVID-19 pandemic in the U.S. In the figure, 100 of the top of keywords used by Trump are linked by co-occurrence.
The graph in Figure~\ref{trump} clusters into communities focused on the following five themes: Republican endorsements (pink), COVID-19 response (green), attacks on news media or Democrats (orange), the economy (dark green), and White House announcements (blue). We will describe more fully how the keyword networks we investigated were formed in the next section.

Word co-occurrence networks have been studied extensively within network science. Network analysis of co-occurrence networks was used to map knowledge structure in scientific fields; see, for example, \cite{kw1,kw2,kw3}. Hashtag co-occurrence networks in Twitter were studied in \cite{h2,ht}. Transcript data of patients with potential Alzheimer’s disease was studied from the context of co-occurrence networks in \cite{mil}. Co-occurrence networks of character names in novels was employed in \cite{bs,rib} to determine communities and key protagonists. Model selection techniques using machine learning were employed in \cite{char} to align co-occurrence networks in movie scripts with various complex network models, such as preferential attachment and random graphs with given expected degree sequences. A detailed review of the literature up to 2014 on word co-occurrence networks may be found in \cite{cong}.

In the present paper, our analysis of Twitter keyword networks does not focus on the meaning of specific tweets, but more on the overarching structure of Twitter keyword networks from political figures over the year 2020. We chose political figures as they tend to generate a consistent number of tweets over time.  We analyzed patterns in the narrative and vocabulary used over the year across hundreds of accounts. A multi-year study of the keyword networks from the tweets of President Trump revealed a low number of communities; see \cite{bl}. On average, and over several years, Trump's tweets clustered into at most five communities.  The results of \cite{bl} and the much larger dataset discussed in Section~\ref{data} led us to hypothesize that tweets organize themselves into a small number of communities for a distinct user. In particular, the hypothesis proposes that users on Twitter post about a small number of topics, using whatever keywords that are relevant to them.

We organize the discussion in this paper as follows. In Section~\ref{secsch}, we consider our framework for the analysis of networks of Twitter keywords. We hypothesize that tweets organize themselves into a low number of communities for a distinct user, and refer to this thesis as the small community hypothesis. Our methods are detailed in Section~\ref{data}, which describes the mining of keyword networks supported by over seven hundred political figures in the U.S.\ and Canada. Our results support the small community hypothesis and are strengthened by considering control data such as random words and tweets formed by AI using GPT-2. We finish with a discussion of our results and propose future work.

We consider undirected graphs with multiple undirected edges throughout the paper. Additional background on graph theory and complex networks may be found in the book~\cite{BonatoCourseWeb}.

\section{Small Community Hypothesis}\label{secsch}

There is no universal consensus on a precise definition for a community in a network. According to Flake, Lawrence, and Giles \cite{Flake_IdentificationWebCommunities2000},  a \emph{community} in a graph is a set of vertices with more links to community members than to non-members. Modularity is defined as
\begin{equation*}\label{ch3eq:modularity}
\frac{1}{2m}\sum_{i,j}\left( A_{i,j}-\frac{k_ik_j}{2m}\right)\delta(c_i,c_j),
\end{equation*}
where $A_{i,j}$ is the weight of the edge between $i$ and $j$, $k_i$ is the sum of the weights of all edges incident to vertex $i$, $c_i$ is the community of which vertex $i$ is a part, $\delta$ is the Kronecker delta function, and $m$ is the sum of all edge weights, given by: $m=\frac{1}{2}\sum_{i,j}A_{i,j}$.

Modularity is used to measure the quality of a partition of a graph and also as an objective function to optimize. Exact modularity optimization is a computationally hard problem, so we look to approximation algorithms to calculate a graph partition. For this, we used the Louvain algorithm \cite{Louvain} that takes a greedy approach to modularity optimization.

As discussed in the introduction, an analysis of keywords taken from Trump's tweets over multiple years revealed a small number of communities. The \emph{small community hypothesis} (or \emph{SCH}) states that the tweets from any user, given sufficient volume, will group themselves into a low number of thematically related clusters. While the SCH does not predict the exact number of clusters, data presented in Section~\ref{data} suggests that in Twitter keyword networks, the number of communities is typically less than ten. For example, the keyword networks in Figures~\ref{trud1} of Prime Minister Justin Trudeau and \ref{trump} of President Trump resolve into four and five communities, respectively.

Note that the SCH does not predict what communities occur in an individual user's Twitter account, or how such communities change over time. Instead, we view it as an emergent, quantitative property of Twitter keyword networks.

In the next section, we will describe our methods and data, which we extended to a much wider dataset of Twitter users. Further, we considered two other datasets generated as control groups to test our methodology.

\section{Methods and Data}\label{data}

Twitter has an Application Programming Interface (or API) that can be accessed for free, with some restrictions. An API is a way for requesting information through a computer program, which we used for retrieving tweets from our users of interest. To take advantage of this API, we used the Tweepy Python library \cite{Tweepy}, which allows ready access to Twitter data with the Python programming language. Occasionally, there was an issue within the code or API, and as a result, nothing was returned. The most notable case of this was for the account of President Trump with handle @realDonaldTrump. For his account, we downloaded the tweets manually from the Trump Twitter Archive (or TTA), a third party website which keeps a record of all of President Trump's tweets, including those that may have been deleted \cite{TrumpTwitterArchive}. This is an especially invaluable resource now that @realDonaldTrump has been suspended from the platform, and it is no longer possible to view his historical Twitter feed on the official site.

For a particular user, the Python code we wrote performed the data processing as follows:
\begin{enumerate}
  \item Collected tweets from Twitter API or Trump Twitter Archive.
  \item Removed retweets and non-English tweets if necessary.
  \item Tokenized tweets, removed stop words, calculated monthly keyword frequency.
  \item For each month, generated $100 \times 100$ adjacency matrices of the top 100 keywords by frequency.
\end{enumerate}

The first step used Tweepy and the Twitter API to gather data between dates. For President Trump, we downloaded the data in CSV format from the TTA, again specifying a range of dates. We obtained a CSV file including all relevant data. This includes the tweet itself, along with metadata including date, number of retweets, and number of likes. In the second step, we filtered out retweets. We only considered tweets in English (otherwise, this would skew the number of communities detected). Additionally, for Canadian and American politicians, typically non-English tweets are just duplicates of English tweets. Prime Minister Trudeau is a good example of that phenomenon, as he tweets everything once in English and a second time in French.
\begin{figure}[!ht]
  \centering
  \includegraphics[width=8cm]{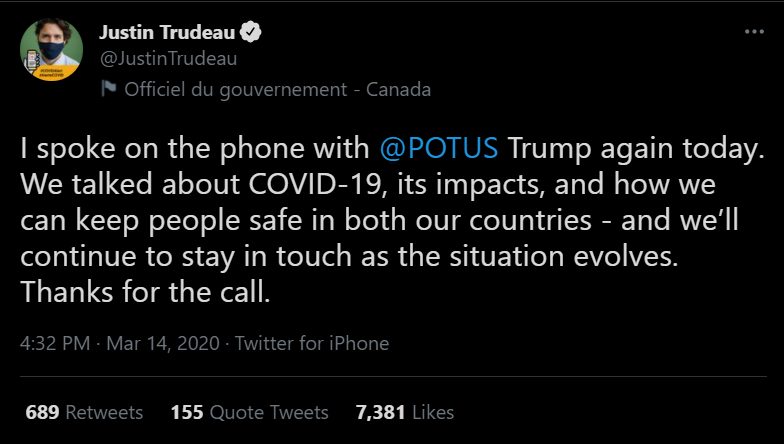}
  \caption{A tweet from the official Twitter account of Prime Minister Trudeau \cite{TrudeauTweetMar142020}.}\label{trudeau}
\end{figure}
The third step involved common NLP techniques to break down the sentences into something more manageable for a machine to process. \emph{Tokenization} is the process of extracting words from a sequence of characters \cite{Chiarcos_LovelyTokens2012}. Once we have identified the words, we can remove the stop words, which are the common and less meaningful words. For our purposes, we used the Natural Language ToolKit Python library, which includes a corpus of stop words \cite{NLTK}. These stop words mostly include articles and pronouns (for example, ``the," and ``he"), and others added manually.

Once the keywords were identified, we calculated the monthly frequency of each word, and kept the top 100 for each month. In the last step, we created the $100 \times 100$ adjacency matrices to represent a graph. For two words $i$ and $j$, and an adjacency matrix $M$, the position $M(i,j)$ contains a numeric value indicating the number of times that both words $i$ and $j$ appeared in the same tweet by the user in the month. This represents the weight of the edge between $i$ and $j$ in the weighted graph described by $M.$ For the tweet in Figure~\ref{trudeau}, we would add 1 to the weight of the edge between every pair of keywords in this tweet, for instance between the vertices for ``covid-19" and ``safe." Recall that we only consider pairs where both words appear in the top 100 keywords by frequency in the month. With steps (1) through (4) completed, data was visualized with Gephi using the ForceAtlas2 (FA2) layout algorithm and Louvain community detection algorithm; see \cite{bast} for more on Gephi.

\begin{figure}[!htpb]
  \centering
  \includegraphics[width=9cm]{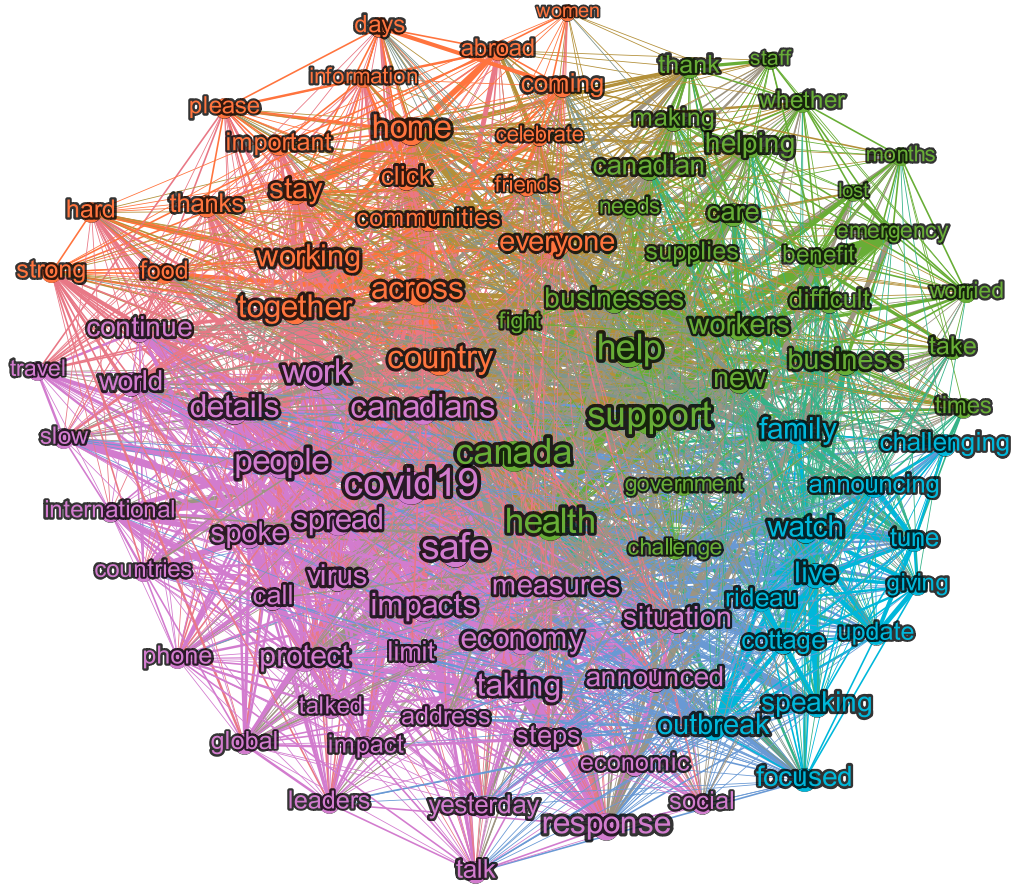}
  \caption{Keyword network created from tweets by Prime Minister Trudeau in March of 2020. }\label{trud1}
\end{figure}

March 2020 marked the beginning of preventative measures taken in North America, such as Ontario closing public schools on March 12th, the day after the WHO declared a global pandemic \cite{CovidTimelineCanada_GlobalApr2020}. The Twitter keyword networks of President Trump and Prime Minister Trudeau are presented in Figure~\ref{trump} and Figure \ref{trud1}, respectively.

The Trump Twitter keyword network had five communities in March 2020, while the one of Trudeau has four; these are the colored sets of nodes in the figures. Similar patterns were detected in the later months of 2020, as will be described in the next section.

\subsection{Political figures}

In an effort to validate the SCH, we formed Twitter keyword networks from 703 political figures from Canada and the U.S.\ From Canada, there were 190 accounts mainly composed of Members of Parliament, with a few other figures from initial testing, such as Ontario Premier Doug Ford. The data from Canadian politicians included 94 Liberals, 71 Conservatives, 16 NDPs, two Green Party members, one from the Saskatchewan Party, and six Independents. From the United States, there were 513 accounts including state governors and members of congress. The data from American politicians included 242 Republicans, 268 Democrats, and three Independents. The accounts of President Trump and Prime Minister Trudeau were also included. The total number of tweets scraped in our analysis was 562,425. We refer the reader to https://cutt.ly/VWtT4bA where we list approximately 12,000 community numbers for the accounts we scraped.

Bilingual accounts (especially among Canadian politicians) posed a challenge since our analysis is based on English tweets only; if these tweets occupied only a minority of the feed, then the rest of the tweets by the author could be kept in the dataset. We also removed retweets, though in some cases this is not ideal since some accounts use this feature as a large share of their communications.

For this dataset, we employed a Python package for community detection with NetworkX~\cite{PythonLouvain} to automate the process described in the previous section. One challenge was the aspect of randomness associated with the Louvain algorithm, where we could run the algorithm twice on identical networks and derive different results. To combat this, for each network, we ran the detection algorithm one hundred times and took the most frequently derived result.

\begin{figure}[!htpb]
  \centering
  \includegraphics[width=9cm]{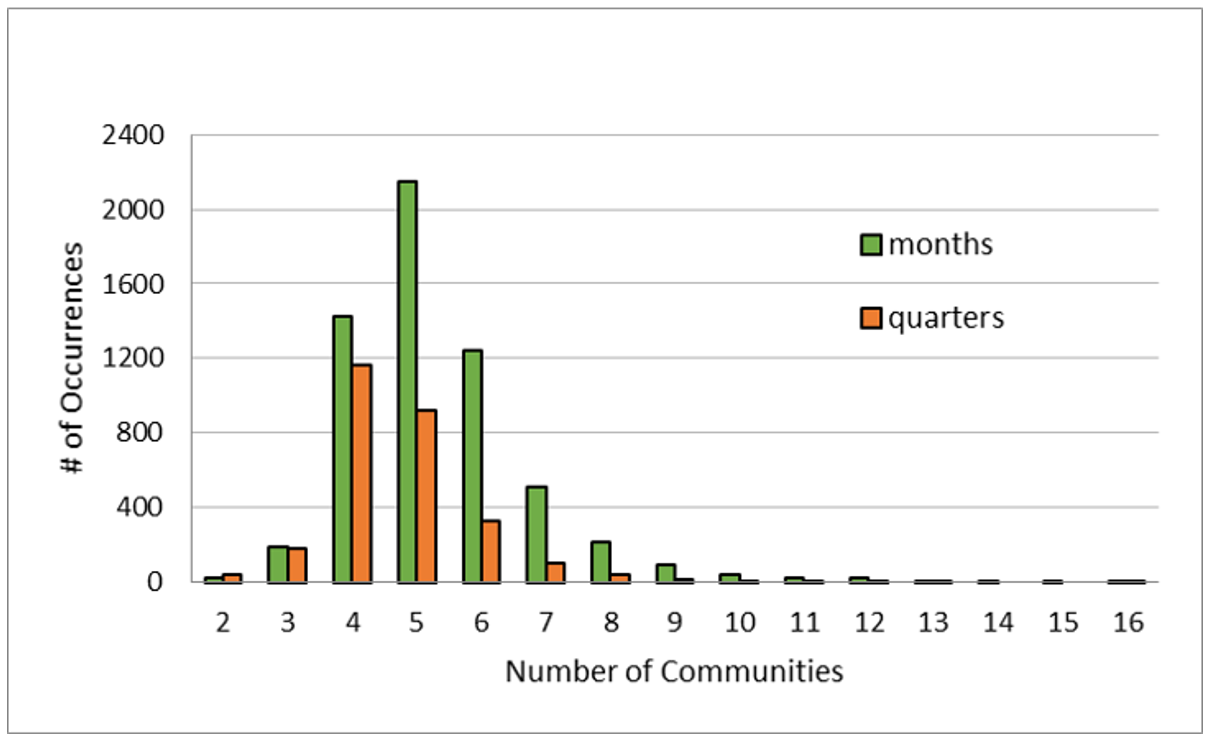}
  \caption{Frequency of number of communities found using the Louvain algorithm on Twitter keyword networks from 703 politicians in 2020, from January to December. Results are shown for monthly and quarterly keyword networks.}\label{pf}
\end{figure}

We analyzed tweets grouped by quarter and by month. See Figure~\ref{pf} for the monthly and quarterly distribution of communities found in the keyword networks of Canadian and American politicians. From the difference in these two sets of graphs, we observed that overall the distribution of data shifted left, with the most frequent number of communities changing from five to four, thus supporting the SCH. This also had the effect of reducing the size of the right tail, with fewer outlying networks containing large community numbers.

\begin{figure}[H]
  \centering
  \subfloat[U.S.\ governors.]{\includegraphics[width=0.5\textwidth]{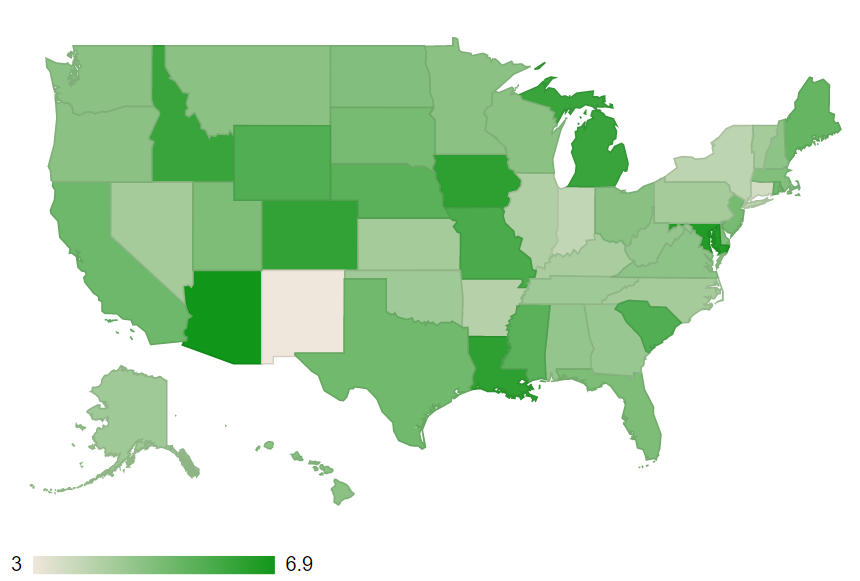}}
  \hfill
  \subfloat[Canadian premiers.]{\includegraphics[width=0.5\textwidth]{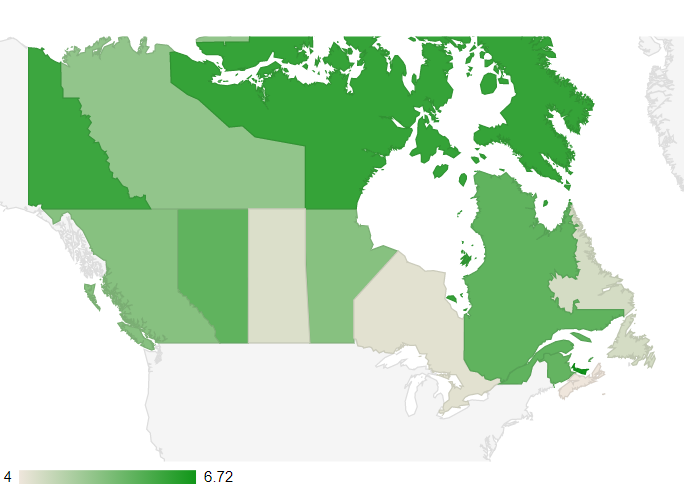}}
  \caption{Heatmap revealing the number of communities based on U.S.\ gubernatorial tweets and those of Canadian premiers. The monthly community counts were averaged over 2020. } \label{hm}
\end{figure}

Overall these findings are in line with our observations from President Trump and Prime Minister Trudeau in Section~\ref{data}, with the relatively low number of communities, centering around four to six. Though the number of communities may appear to decrease with a higher volume of tweets (see quarterly versus monthly numbers), we did not find any significant correlation between these variables. We observed that if there were a relatively low number of tweets in a month, the network was more likely to contain a larger number of communities. A heat map recording the number of communities using gubernatorial tweets and those of premiers may be found in Figure~\ref{hm}.

\begin{figure}[!htpb]
  \centering
  \includegraphics[width=7cm]{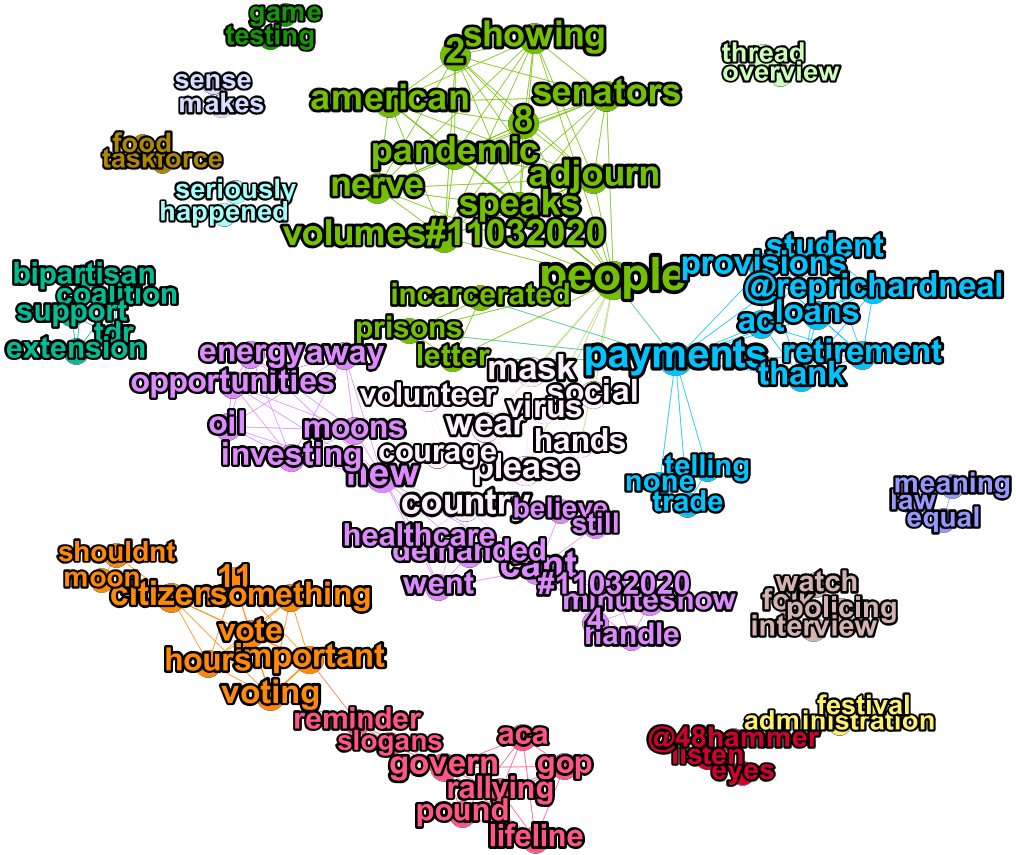}
  \caption{Keyword networks for October 2020 of U.S.\ House Representative Danny K.\ Davis of Illinois' 7th congressional district. The data was made up of 35 tweets, and 16 communities were detected.} \label{dan}
\end{figure}

In the case of sparse data, the resulting network was often disconnected, which lead to unpredictable results. In particular, the Louvain algorithm occasionally assigned small connected components, such as those associated with individual tweets, to their communities. For example, U.S.\ Representative Danny K.\ Davis of Illinois had only 35 tweets written in October 2020, which lead to 16 communities detected by the Louvain algorithm; see Figure~\ref{dan}.

\subsection{Pseudo-tweets}

\begin{figure}[!htpb]
  \centering
  \includegraphics[width=8cm]{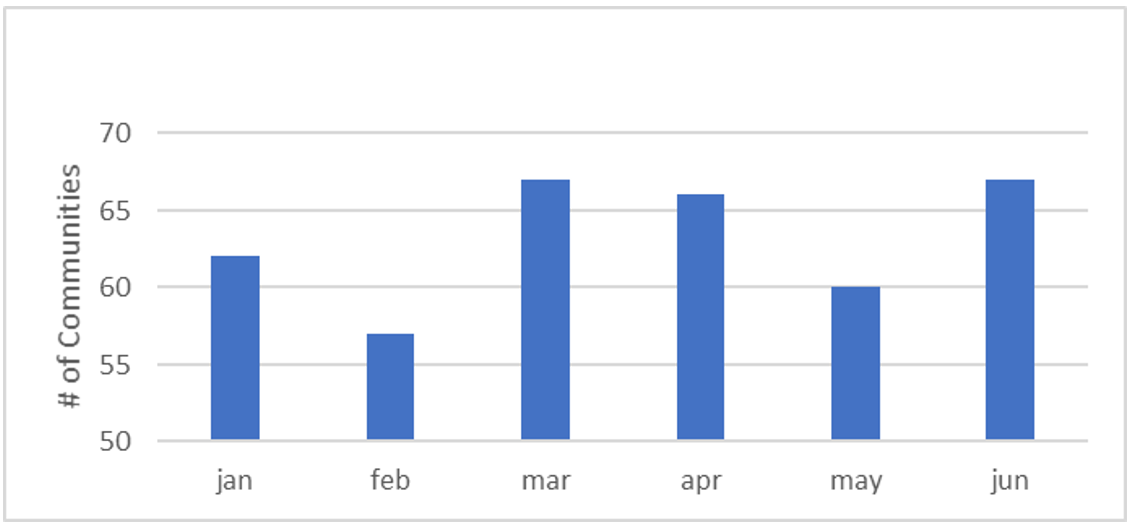}
  \caption{Number of communities found each ``month" using pseudo-tweets composed of random English words.}\label{com}
\end{figure}

To further test the validity of the SCH, we generated other related datasets as control groups. These other datasets consist of what we refer to as \emph{pseudo-tweets} that did not come from an account on Twitter. The goal of the analysis of this data was to detect what phenomena influence the small community hypothesis. In particular, the SCH may be influenced by the vocabulary of a specific user, or how language is implemented. The first dataset, and perhaps most obvious, is one composed of randomly generated tweets. These are simply a set of random English words adhering to the restriction on length of tweets (that is, 280 characters). The expected results from analysis of this dataset was a large number of communities, or no pattern at all, due to the pseudo-tweets not using language as a human would with regards to word choice and sentence structure. The method for generating these messages does not weight word choice by any measure of popularity, nor does it adhere to grammar rules. We generated six ``months" worth of data with 100 tweets per month and ran the community detection on the results. See Figure \ref{com} for the resulting number of communities from each month. The algorithm revealed 57 to 67 communities, which is certainly not small when compared to the number of communities found in politicians.

\begin{figure}[H]
  \centering
  \includegraphics[width=8cm]{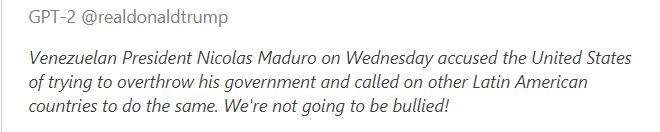}
  \caption{A sample pseudo-tweet generated using the public GPT-2 model, fine tuned on real tweets from @realDonaldTrump.}\label{gpt}
\end{figure}

\begin{figure}[h!]
  \centering
  \includegraphics[width=8cm]{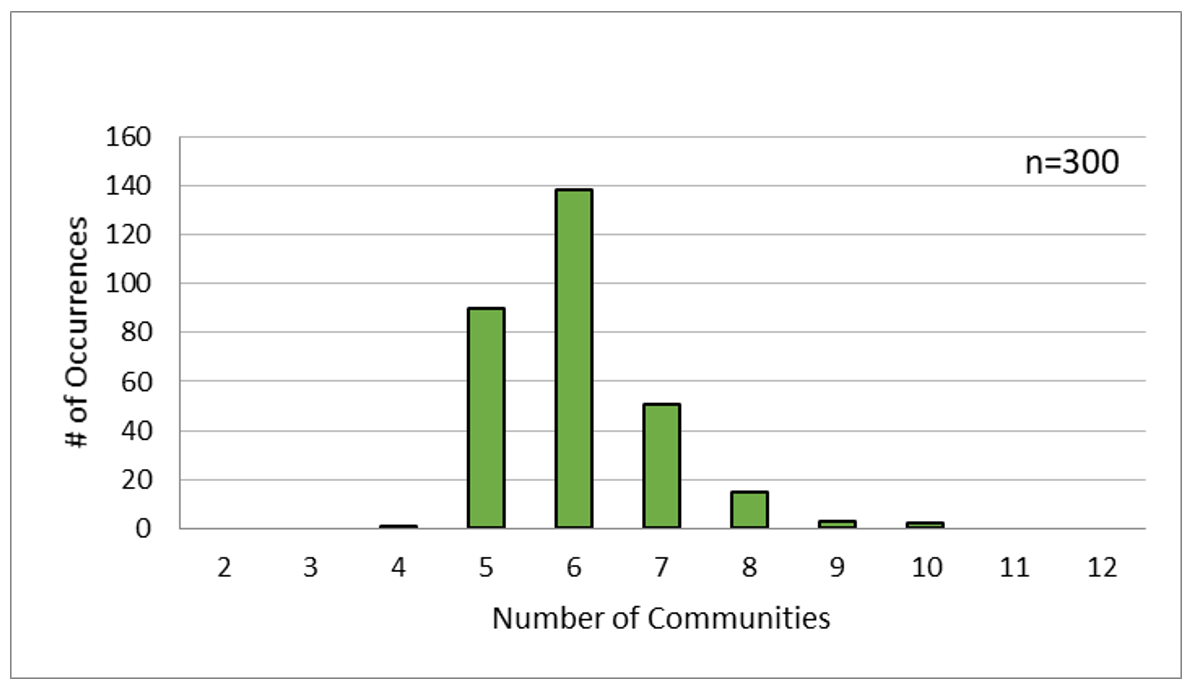}
  \caption{Frequency of communities found in GPT-2 generated keyword networks.}\label{hist}
\end{figure}

Another dataset we considered was one containing computer-generated tweets. In recent years, there has been a rapid improvement in the quality of linguistic algorithms mimicking human speech. A well-known example of this is the use of deep learning to create nearly indistinguishable fake videos of celebrities and politicians, also known as \emph{deepfakes} \cite{Guardian_Deepfakes}. A related application of deep learning is an advanced text-based language engine developed by OpenAI. GPT-2 is the second generation of an unsupervised learning algorithm which has been pre-trained on 40GB of internet data and will predict the next word given a piece of text \cite{GPT2}. Due to the authors' concerns about malicious application of the full algorithm, only a smaller version is available publicly for research and testing. Using a Python implementation of GPT-2 in \cite{GPT2Woolf}, we generated a large number of pseudo-tweets. These were limited by character length, though a bit more roughly since we wanted to allow the AI to finish its sentences. We used the model to generate tweets mimicking those of President Trump by feeding in all of the @realDonaldTrump tweets we had gathered in the initial analysis, using data from January to June of 2020. See Figure~\ref{gpt} for a sample pseudo-tweet generated by GPT-2. We observed that the messages were generally quite coherent and, in our view, their content was in line with President Trump's preferred vocabulary.

We ran the model 50 times, each one creating six months of 100 tweets each. That is, there were 300 networks and 30,000 total pseudo-tweets.  We then applied our community analysis algorithms; see Figure~\ref{hist} for the resulting frequency of communities. Compared to the random English words from the last analysis, this distribution is more in line with what we expect from a human-authored Twitter account.

\section{Discussion and Future Work}

Twitter is one of the most popular social networks owing in part to its accessibility and short format. The restriction on message length creates an information-dense medium that lends itself well to a networked keyword analysis. Using community detection and other network science tools, we analyzed this rich data source as a set of complex networks of keywords.

We proposed the small community hypothesis, where Twitter keyword networks cluster into a low number of communities. We tested the hypothesis on data scraped from Twitter and comprised of tweets of politicians in the U.S.\ and Canada. The data set contained 562,425 tweets from 703 accounts for all of 2020. From the results of the Louvain community algorithm, we found that over 75\% of months fell between four and six communities across all accounts. Our results suggest that the SCH is an observable phenomenon within Twitter keyword networks. We also tested the hypothesis on two other datasets, one of random English words and another of pseudo-tweets generated by the GPT-2 deep learning model. The former gave large community numbers, and the latter dataset gave results closer to the original data.

One direction for future work would be to probe the origins of SCH and whether it is a random occurrence within Twitter keyword networks or, for an example, a consequence of how humans use language. The SCH may be foundational in how users approach social media, focusing their messaging on a small number of topics. The fact that the randomized pseudo-tweet datasets had a much larger number of communities compared to those generated by GPT-2 (which more closely follows actual tweets) indicates that the SCH could be an artefact of language.

Another direction would be to expand our analysis to Twitter accounts of non-political figures. These may include journalists, actors, or accounts of public figures that tweet with enough volume. We had hoped to include more data from the accounts of prominent public figures outside of the political sphere, but we encountered problems with the API and with the code for data processing, mostly due to sporadic user inactivity. We did examine 24 Twitter accounts of public figures who are not politicians but who tweet often, and found analogous results with on average seven communities monthly across 2020.

\section{Acknowledgments} The research for this paper was supported by grants from NSERC and Ryerson University.


\begin{thebibliography}{99}

\bibitem{bast} M.\ Bastian, S.\ Heymann, M.\ Jacomy, Gephi: an open source software for exploring and manipulating networks,
In: \emph{Proceedings of the International AAAI Conference on Weblogs and Social Media}, 2009.

\bibitem{bs} A.\ Beveridge, J.\ Shan, Network of Thrones, \emph{Math Horizons Magazine} \textbf{23} (2016) 18--22.

\bibitem{Louvain}
V.\ Blondel, J.L.\ Guillaume, R.\ Lambiotte, E.\ Lefebvre, Fast unfolding of communities in large networks, \emph{Journal of Statistical Mechanics: Theory and Experiment} \textbf{10} (2008).

\bibitem{BonatoCourseWeb}
A.\ Bonato, \emph{A Course on the Web Graph}, American Mathematical Society Graduate Studies Series in Mathematics, Providence, Rhode Island, 2008.

\bibitem{char} A.\ Bonato, D.R.\ D'Angelo, E.R.\ Elenberg, D.F.\ Gleich, Y.\ Hou, Mining and modeling character networks, In: \emph{Proceedings of the 13th Workshop on Algorithms and Models for the Web Graph}, 2016.

\bibitem{bl} A.\ Bonato, L.\ Roach, The math behind Trump’s tweets, \emph{The Conversation}, 2018.

\bibitem{Chiarcos_LovelyTokens2012}
C.\ Chiarcos, J.\ Ritz, M.\ Stede, By all these lovely tokens... Merging conflicting tokenizations, \emph{Lang Resources \& Evaluation} \textbf{46} (2012) 53–74.

\bibitem{PythonLouvain}
Community detection for NetworkX. Accessed September 1, 2021 from \url{https://python-louvain.readthedocs.io/en/latest/}.

\bibitem{cong} J.\ Cong, H.\ Liu, Approaching human language with complex networks, \emph{Physics of Life Reviews} \textbf{11} (2014) 598--618.

\bibitem{Flake_IdentificationWebCommunities2000}
G.W.\ Flake, S.\ Lawrence, C.L\. Giles, Efficient identification of Web communities, In: \emph{Proceedings of the Sixth ACM-SIGKDD International Conference on Knowledge Discovery and Data Mining}, 2000.

\bibitem{kw1} J.\ Law, S.\ Bauin, J.\ Courtial, J.\ Whittaker, Policy and the mapping of scientific change: A co-word analysis of research into environmental acidification. \emph{Scientometrics} \textbf{14} (1988) 251--264.

\bibitem{mil} T.\ Millington, S.\ Luz, Analysis and classification of word co-occurrence networks from Alzheimer’s patients and controls, \emph{Front. Comput. Sci.} \textbf{3}:649508, 2021.

\bibitem{NLTK}
Natural Language Toolkit. Accessed September 1, 2021 from \url{https://www.nltk.org/}.

\bibitem{GPT2}
A.\ Radford, J.\ Wu, R.\ Child, D.\ Luan, D.\ Amodei, I.\ Sutskever, Language Models are unsupervised multitask learners, \emph{OpenAI}, 14 Feb 2019. Accessed on September 1, 2021 from \url{https://openai.com/blog/better-language-models/}.

\bibitem{kw2} S.\ Radhakrishnan, S.\ Erbis, J.A.\ Isaacs,  S.\ Kamarthi, Novel keyword co-occurrence network-based methods to foster systematic reviews of scientific literature, \emph{PLOS ONE} \textbf{12}(3): e0172778, 2017.

\bibitem{rib} M.A.\ Ribeiro, R.A.\ Vosgerau, M.L.P.\ Andruchiw, S.\ Ely de Souza Pinto, The complex social network from
the Lord of the Rings, \emph{Rev. Bras. Ensino Fs} \textbf{38} (2016) 1304

\bibitem{Guardian_Deepfakes}
I.\ Sample, What are deepfakes---and how can you spot them?, \emph{The Guardian}, 13 Jan 2020. Accessed September 1 from \url{https://www.theguardian.com/technology/2020/jan/13/what-are-deepfakes-and-how-can-you-spot-them}.

\bibitem{CovidTimelineCanada_GlobalApr2020}
Timeline: How canada has changed since coronavirus was declared a pandemic, \emph{Global News}, 11 Apr. 2020. Accessed September 1, 2021 from \url{https://globalnews.ca/news/6800118/pandemic-one-month-timeline/}.

\bibitem{TrumpTwitterArchive}
Trump Twitter Archive V2. Accessed September 1 from \url{https://www.thetrumparchive.com/}.

\bibitem{Tweepy}
Tweepy Documentation. Accessed September 1 from \url{https://docs.tweepy.org/en/latest/api.html}.

\bibitem{TrudeauTweetMar142020}
Twitter account of Justin Trudeau ``I spoke on the phone with @POTUS Trump again today...". Accessed September 1, 2021 from \url{https://twitter.com/JustinTrudeau/status/1238926089637986306}.

\bibitem{h2} R.\ Wang, W.\ Liu, S.\ Gao, Hashtags and information virality in networked social movement, \emph{Online Information Review} \textbf{40} (2016) 850--866.

\bibitem{ht} L.\ Weng, F.\ Menczer, Topicality and impact in social media: diverse messages, focused messengers, \emph{PLOS ONE} \textbf{10} e0118410, 2015.

\bibitem{GPT2Woolf}
M.\ Woolf, \emph{How To Make Custom AI-Generated Text With GPT-2}, 4 Sep 2019. Accessed September 1, 2020 from \url{https://minimaxir.com/2019/09/howto-gpt2/}

\bibitem{kw3} T.\ You, J.\ Yoon, O,\ Kwon, W.\  Jung, Tracing the evolution of physics with a keyword co-occurrence network, \emph{Journal of the Korean Physical Society} \textbf{78} (2021) 236--243.

\end{thebibliography}
\end{document}